\documentclass[12pt]{elsarticle}

\usepackage{amsmath}
\usepackage{amssymb}
\usepackage{color}
\usepackage{tabularx}
\usepackage{mathtools}
\usepackage{algpseudocode}
\usepackage{enumitem}
\usepackage{multirow}
\usepackage{blindtext}
\usepackage{epstopdf}  
\usepackage{hyperref}
\usepackage{bm}
\usepackage{ragged2e}
\usepackage{natbib}
\biboptions{sort&compress}

\usepackage{etoolbox}
\makeatletter
\patchcmd{\ps@pprintTitle}{\footnotesize\itshape
       Preprint submitted to \ifx\@journal\@empty Elsevier
       \else\@journal\fi\hfill\today}{\relax}{}{}
\makeatother

\newcommand{\br}{\bm{r}}
\newcommand{\bR}{\bm{R}}

\newcommand{\ba}{\bm{a}}
\newcommand{\btheta}{\bm{\theta}}

\begin{document}

\begin{frontmatter}
\title{
Solving Many-Electron Schr\"odinger Equation Using Deep Neural Networks
}
\author[address1]{Jiequn Han}
\ead{jiequnh@princeton.edu}
\author[address1]{Linfeng Zhang\corref{mycorrespondingauthor}}
\cortext[mycorrespondingauthor]{Corresponding author}
\ead{linfengz@princeton.edu}
\author[address1,address2,address3]{Weinan E}
\ead{weinan@math.princeton.edu}

\address[address1]{Program in Applied and Computational Mathematics, 
Princeton University, Princeton, NJ 08544, USA}
\address[address2]{Department of Mathematics, 
Princeton University, Princeton, NJ 08544, USA}
\address[address3]{Beijing  Institute of Big Data Research, 
Beijing, 100871, P.R.~China}

\begin{abstract}
We introduce a new family of trial wave-functions based on deep neural networks to solve the many-electron Schr\"odinger equation. The Pauli exclusion principle is dealt with explicitly to ensure that  the trial wave-functions are physical. 
The optimal trial wave-function is obtained through variational Monte Carlo and the computational cost scales quadratically with the number of electrons.
The algorithm does not make use of  any prior knowledge such as atomic orbitals.
Yet it is able to represent accurately the ground-states of the tested systems, including He, H${}_2$, Be, B, LiH, and a chain of 10 hydrogen atoms. 
This opens up new possibilities for solving large-scale many-electron Schr\"odinger equation.
\end{abstract}

\begin{keyword}
Schr\"odinger equation \sep Variational Monte Carlo \sep Deep neural networks \sep Trial wave-function
\end{keyword}

\end{frontmatter}

\section{Introduction}
An accurate quantum mechanical treatment of the interaction between many electrons and ions is the foundation for  modeling physical, chemical, and biological systems. 
Theoretically, these systems are described by the many-electron Schr\"odinger equation, which consists of the kinetic  and Coulomb interaction terms  \cite{dirac1929quantum}.
Solving  the Schr\"odinger equation accurately for real physical systems has been prohibitively difficult due to the high dimensionality of the  Hilbert space involved,
the high degree of entanglement produced by the electron-electron and electron-ion interactions,
and the notorious Pauli exclusion principle imposed on the wave-function, i.e., the wave-function has to change its sign when two identical electrons exchange places~\cite{pauli1925principle}.

Developing efficient algorithms for this problem is among the most heroic endeavors in computational science and  has achieved  remarkable successes. 
An incomplete list  of the major methodologies developed so far  includes 
the Hartree-Fock (HF) based methods \cite{roothaan1960RHF,pople1954UHF},
configuration interaction (CI) based methods \cite{pople1987QCI,werner1988MRCI,knowles1988MRCI,shiozaki2011MRCI-F12}, 
coupled cluster (CC) based schemes \cite{purvis1982CCSD,paldus1999CC,bartlett2007CC,shavitt2009CC},  
Monte Carlo-based approaches \cite{mcmillan1965VMC,ceperley1977VMC,bressanini1999vmc,blankenbecler1981AFQMC,zhang1997AFQMC,zhang2003QMC,van2010diaMC,foulkes2001QMC}, 
and the more recently developed density matrix renormalization group (DMRG) theory \cite{white1992DMRG,white1999DMRG,chan2002DMRG,chan2016DMRG,stoudenmire2017slicedDMRG}
and density matrix embedding theory (DMET) \cite{knizia2012DMET,knizia2013DMET,wouters2016practicalDMET}. 
We refer to Refs.~\cite{grotendorst2000modernQC,szabo2012modernQC,motta2017towards} for a more detailed review of these advances.

Of particular interest to this work is the variational Monte Carlo (VMC) scheme which uses  variational principle and Monte Carlo sampling to obtain the best parametrized trial wave-function~\cite{mcmillan1965VMC,ceperley1977VMC,bressanini1999vmc,foulkes2001QMC}.
Naturally the key component is the representation of the trial wave-functions. 
The most commonly used trial wave-functions typically consist of an anti-symmetric Slater determinant~\cite{slater1930note} multiplied by a symmetric Jastrow correlation factor~\cite{jastrow1955many}.
There have been tremendous efforts  on improving the nodal surface (a subspace on which the function value equals zero and across which it changes the sign) of the anti-symmetric part of the trial wave-function and  the representability of the symmetric part~\cite{umrigar1988optimized,umrigar2007alleviation,casula2003geminal,changlani2009CPS}.

With remarkable advances in many fields such as computer vision and speech recognition, the deep neural network (DNN) has shown great capacity in approximating high-dimensional functions 
(see, e.g., review \cite{lecun2015deep} and the references therein). 
Furthermore, DNN has been successfully used in solving general high-dimensional partial differential equations~\cite{han2018solving,e2017deep,berg2017unified,khoo2017solving}
and certain quantum many-body problems for Bosonic and lattice systems~\cite{saito2018method,carleo2017solving,gao2017QMB,saito2017QMB,cai2018QMB}. 
However, there have been few attempts to solve the many-electron Schr\"odinger equations based on DNN, and this constitutes the main objective of this work.

To achieve this, we develop a general and efficient DNN representation for the many-electron wave-function satisfying the Pauli exclusion principle.
The resulted trial wave-function can naturally fit into the framework of VMC to optimize the parameters in our model.
As preliminary tests, we show that this DNN-based trial wave-function is able to produce reasonably well ground-state energies for some small systems, such as Be, B, LiH, and a chain of 10 hydrogen atoms (H${}_{10}$).
In addition, learning from scratch without any prior knowledge and without resorting to a reference of atomic bases, 
the DNN-based wave-function is able to reproduce the electronic structures of the tested systems.
We call the methodology introduced here the Deep WaveFunction method, abbreviated DeepWF.
This paper only reports our initial results.  There is still a huge room for improvement.  

\section{Method}
\begin{figure*}
  \centering
  \includegraphics [width=0.98\textwidth] {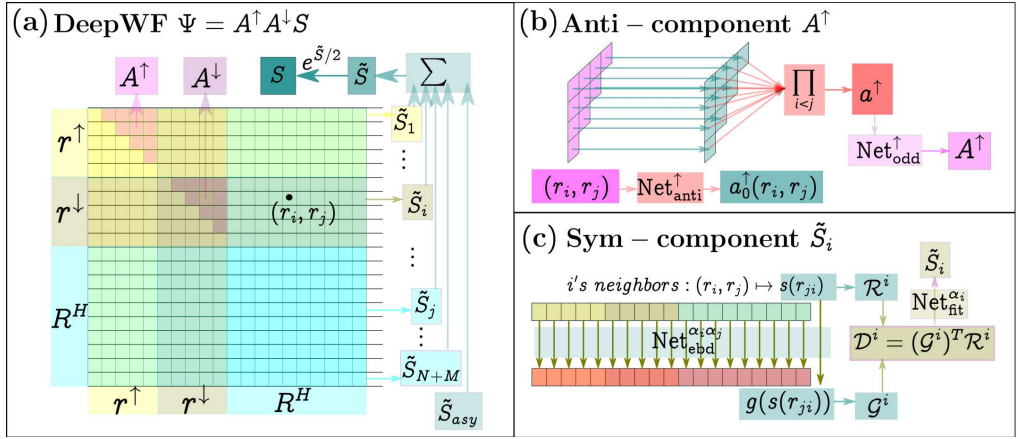}
  \caption{(color online). Schematic plot of DeepWF, taking H${}_{10}$ as an example. 
$\bm{r}^{\uparrow}$ and $\bm{r}^{\downarrow}$ denote the positions of 5 electrons of spin-up and 5 of spin-down, respectively, and $\bR^H$ denotes the positions of 10 hydrogen atoms.  
  (a) the entire wave-function $\Psi(\br;\bR)=S(\br;\bR)A^\uparrow(\br^\uparrow)A^\downarrow(\br^\downarrow)$, decomposed into symmetric function $S(\br;\bR)$ and two anti-symmetric functions $A^\uparrow(\br^\uparrow), A^\downarrow(\br^\downarrow)$, with respect to spin-up and spin-down electrons; 
  (b) anti-symmetric function $A^\uparrow(\br^\uparrow)$ for spin-up electrons; 
  (c) symmetric function $\tilde{S}_i(\mathcal{R}^i)$ as the contribution of particle $i$ in the symmetric function $\log(S^2(\br;\bR))$. See text for details.}
  \label{fig:scheme}
\end{figure*}

We consider a system  of $N$ electrons and $M$ ions, under the Born-Oppenheimer approximation~\cite{Born1927BO}. 
This system is described by the Hamiltonian
\begin{align}
\hat{H}= &-\frac12\sum_{i=1}^N\nabla_i^2 + \sum_{i=1}^N\sum_{j=i+1}^N\frac{1}{|\br_i-\br_j|} \nonumber \\
&-\sum_{I=1}^M\sum_{i=1}^N\frac{Z_I}{|\br_i-\bR_I|} + \sum_{I=1}^M\sum_{J=I+1}^M\frac{Z_IZ_J}{|\bR_I-\bR_J|},
\end{align}
where $\br=(\br_1,\dots,\br_N)$ and $\bR=(\bR_1,\dots,\bR_M)$ are the coordinates of the electrons and the ions, respectively, and $Z_I$ denotes the nuclear charge. 
Since $\hat{H}$ is spin-independent, it is valid to assume that the first $N_{\uparrow}$ electrons are of spin-up and the remaining $N_{\downarrow}=N-N_{\uparrow}$ electrons are of spin-down~\cite{foulkes2001QMC}. 
Accordingly, we can write the wave-function in a spin-independent form $\Psi(\br;\bR)$. 
Let $N_{\text{tp}}$ be the number of ion types in the system, then there are $N_{\text{tp}}+2$ types of particles in total, taking into account the electrons of spin-up and spin-down separately. 

In this work we restrict our attention to the ground state of the system. 
The variational principle states that the wave-function associated with the ground state minimizes within the required symmetry the following energy functional 
\begin{align*}
E[\Psi] =
\frac{\int\Psi^{*}(\br;\bR)\hat{H}\Psi(\br;\bR)\,d\br}{\int\Psi^{*}(\br;\bR)\Psi(\br;\bR)\,d\br}.
\end{align*}

We now discuss how to represent the wave-function $\Psi(\br;\bR)$ with DNN. The trial wave-functions are assumed to be real-valued.
The primary goal is to ensure that the represented wave-function satisfies the anti-symmetry property.
To this end, we decompose the wave-function as $$\Psi(\br;\bR)=S(\br;\bR)A^\uparrow(\br^\uparrow)A^\downarrow(\br^\downarrow),$$ 
where $\br^\uparrow,\br^\downarrow$ denote the positions of spin-up and spin-down electrons, respectively. We require $S(\br,\bR)$ to be symmetric
and $A^\uparrow(\br^\uparrow), A^\downarrow(\br^\downarrow)$ to be both anti-symmetric.
As an analogy, one can view $S(\br;\bR)$ as a Jastrow factor and $A^\uparrow(\br^\uparrow), A^\downarrow(\br^\downarrow)$ as Slater determinant-like functions. 
In practice, $\log(\Psi^2(\br;\bR))$ 
is more convenient for implementing VMC.
From this perspective, the introduced decomposition of wave-function becomes
\begin{align}
&\log(\Psi^2(\br;\bR)) \nonumber \\
= 
& \tilde{S}(\br;\bR) + \log(|A^\uparrow(\br^\uparrow)|^2) + \log(|A^\downarrow(\br^\downarrow)|^2),
\end{align}
where $\tilde{S}(\br;\bR) \coloneqq \log(S^2(\br;\bR))$ is still a general symmetric function.
On the other hand, although being  symmetric, $\log(|A^\uparrow(\br^\uparrow)|^2)$ and $\log(|A^\downarrow(\br^\downarrow)|^2)$ have singularities on the nodal surface. 
Therefore we need to secure this property by representing $A^\uparrow(\br^\uparrow)$ and $ A^\downarrow(\br^\downarrow)$ directly.
In the following we discuss how to represent $A^\uparrow(\br^\uparrow), A^\downarrow(\br^\downarrow)$, and $\tilde{S}(\br;\bR)$ with DNN and how to learn the optimal parameters using VMC.
See Fig.~\ref{fig:scheme} for a schematic illustration.

The basic building block for the anti-symmetric function $A^\uparrow(\br^\uparrow)$ is an ansatz
\begin{equation}
\label{eq:anti_product}
\ba^\uparrow(\br^\uparrow)=\prod_{1\leq i< j \leq N_{\uparrow}}\ba^\uparrow_{0}(\br_i,\br_j),
\end{equation}
where $\ba^\uparrow_0$ is a general two-body anti-symmetric function with $M_1$-dimensional output, i.e., $\ba^\uparrow_0(\br_i,\br_j)=-\ba^\uparrow_0(\br_j,\br_i)$,
and the products are component-wise. 
In \eqref{eq:anti_product} we have assumed $N_{\uparrow}\geq 2$. 
Such representation can be viewed as a generalization of the Laughlin wave-function~\cite{laughlin1983anomalous}, which describes well the anomalous quantum Hall effect.
A related model is the so-called correlator product
state (CPS)~\cite{changlani2009CPS}, which works well for lattice systems and has natural connections with some recently proposed neural-network quantum states~\cite{glasser2018neural,clark2018unifying}.
In practice, we let 
$\ba^\uparrow_0(\br_i,\br_j)=
\text{Net}^\uparrow_{\text{anti}}(\br_i,\br_j,r_{ji})-\text{Net}^\uparrow_{\text{anti}}(\br_j,\br_i,r_{ji})$, where $\text{Net}^\uparrow_{\text{anti}}(\cdot)$ is represented by a DNN.
By convention $r_{ji}=|\br_{ji}|=|\br_j-\br_i|$ denotes the Euclidean distance between particles $i$ and $j$. Next $\ba^\uparrow$ is fed into another DNN $\text{Net}^\uparrow_{\text{odd}}$,
which outputs a scalar and satisfies $\text{Net}^\uparrow_{\text{odd}}(\bm{x})=-\text{Net}^\uparrow_{\text{odd}}(-\bm{x})$.
We can further adjust its scale and define $A^\uparrow(\br^\uparrow)$ through $\log|A^\uparrow(\br^\uparrow)|=f_{\text{odd}} \log|\text{Net}^\uparrow_{\text{odd}}(\ba^\uparrow(\br))|$ where $f_\text{odd}$ is a positive scalar factor. The sign of $A^\uparrow(\br^\uparrow)$ is taken as the same as $\text{Net}^\uparrow_{\text{odd}}(\ba^\uparrow(\br^\uparrow))$ such that $A^\uparrow(\br^\uparrow)$ inherits the anti-symmetry.
The same construction introduced above are applied to spin-down electrons and yield $A^\downarrow(\br^\downarrow)$.

While DNN-based anti-symmetric function has seldom been investigated in the literature, 
DNN-based symmetric function has gained wide attentions recently in modeling many-body potential energy surface~\cite{behler2007generalized,han2017deep,zhang2018deep,zhang2018end,schutt2017schnet}, 
free energy surface~\cite{zhang2018reinforced,zhang2018deepcg,schneider2017stochastic}, etc.  
For instance, the Deep Potential-Smooth Edition (DeepPot-SE) model is able to efficiently describe the interatomic potential energy of both finite and extended systems
~\cite{zhang2018end}. 
A crucial step there is to faithfully map the input atomic coordinates onto a  symmetry-preserving feature space. 
Therefore, it shares the symmetric property of the Jastrow factor, and might become a more general Jastrow factor that is able to better capture the many-body correlations between the electrons and ions.

We extend the DeepPot-SE model to construct $\tilde{S}(\br;\bR)$. 
For later convenience, we regard $\bR$ as $(\br_{N+1},\dots,\br_{N+M})$ and let $\bar{\br}\coloneqq (\br_1,\cdots,\br_{N+M})$.
For each particle $i$, we consider its neighboring environment in terms of the Cartesian coordinates:
\begin{align}
\label{eq:local_env}
\mathcal{R}^i=\{\bm{s}(\bm{r}_{1i})^T, \cdots, \bm{s}(\bm{r}_{ji})^T, \cdots
\}^T \in \mathbb{R}^{(N+M-1)\times 4}, 
\end{align}
where $\bm{s}(\bm{r}_{ji})=(\br_{ji}/r_{ji}^2, 1/r_{ji})\in \mathbb{R}^4$,
$j\in\{1,\dots,N+M\}\backslash\{i\}$,
are extended relative coordinates. 
Then we decompose $\tilde{S}(\br;\bR)$ into the sum of contributions from each particle and an extra asymptotic term, i.e., 
\begin{align}
\tilde{S}(\br;\bR)= \tilde{S}_{\text{asy}}(\bar{\br}) +\sum_{i=1}^{N+M}\tilde{S}_i(\mathcal{R}^i).
\end{align}

We define $\tilde{S}_i$ in a 
way that guarantees the symmetric property.
We first construct a series of embedding networks $\text{Net}_{\text{ebd}}^{\alpha_i\alpha_j}$,  each mapping the extended relative coordinates $\bm{s}(\br_{ji})$ to $M_2$ components. 
Here $\alpha_i$ denotes the type of particle $i$, 
and the parameters of $\text{Net}_{\text{ebd}}^{\alpha_i\alpha_j}$ depend on the types of both particle $i$ and its neighbor particle $j$.
The embedding matrix $\mathcal{G}^{i}\in\mathbb{R}^{(N+M-1)\times M_2}$ and the encoded feature matrix $\mathcal{D}^i\in\mathbb{R}^{M_2\times 4}$ of particle $i$ are then defined as
\begin{align}
&(\mathcal{G}^{i})_{jk}=(\text{Net}_{\text{ebd}}^{\alpha_i\alpha_j}(\bm{s}(\br_{ji})))_k,\\
&\mathcal{D}^i=(\mathcal{G}^{i})^T\mathcal{R}^i.
\end{align}
By definition one can check that the feature matrix $\mathcal{D}^i$ is invariant under the permutation of particles in the same type, and its {$M_2\times 4$} components are finally reshaped into a vector to serve as the input of a fitting network $\text{Net}^{\alpha_i}_{\text{fit}}$ and yield  $\tilde{S}_i$. 

The remaining term $\tilde{S}_{\text{asy}}(\bar{\br})$ should also be symmetric and serve the additional purposes of making the trial wave-function close to some desired asymptotic properties.
This term has the form
\begin{align}
\label{eq:asymptotic_term}
\tilde{S}_{\text{asy}}(\bar{\br}) = &-\sum_{k=N+1}^{N+M} Z_{k-N}\sum_{i=1}^N \frac{r_{ik}+f_{\text{dec}}r_{ik}^2}{r_{ik} + 0.5} \nonumber \\
&+ \sum_{i=1}^N\sum_{j=i+1}^N \frac{r_{ij}}{r_{ij} + 1},
\end{align}
where $f_{\text{dec}}$ is a positive scalar as decay factor. 
The first purpose is to ensure that $\Psi(\br;\bR)$ is square integrable when $f_{\text{dec}}$ is large enough such that $\Psi(\br;\bR)$ has exponential decay for the electrons that are far away. 
The second purpose is related to the so-called \emph{cusp} conditions, meaning that the trial wave-function has the correct singular behavior when two particles are close to each other~\cite{kato1957cusp}. 
If we let $\log(\Psi^2(\br;\bR)) = \tilde{S}_{\text{asy}}(\bar{\br})$, it can be shown from \eqref{eq:asymptotic_term} that for $1\leq i<j\leq N,~ N+1\leq k\leq N+M$,
\begin{align}
\label{eq:cusp_condition}
\frac{1}{\Psi}\frac{\partial \Psi}{\partial r_{ik}}\bigg|_{r_{ik}=0}=-Z_{k-N}, \quad
\frac{1}{\Psi}\frac{\partial \Psi}{\partial r_{ij}}\bigg|_{r_{ij}=0}=\frac12.
\end{align}
This indicates $\Psi(\br;\bR)$ meets the cusp conditions of both the electron-nuclear pairs and unlike spin pairs (see e.g.,~\cite{kato1957cusp,foulkes2001QMC} for more details).
In numerical computation we find that adding $\tilde{S}_{\text{asy}}(\bar{\br})$  makes the training process much more stable, even though the cusp conditions are not strictly satisfied due to the influence of $\tilde{S}_i$ and $A(\br)$ on the terms in \eqref{eq:cusp_condition}. 
 
We denote all the parameters in the DNNs ($\text{Net}^\uparrow_{\text{anti}}$, $\text{Net}^\downarrow_{\text{anti}}$, $\text{Net}^\uparrow_{\text{odd}}$, $\text{Net}^\downarrow_{\text{odd}}$, $\text{Net}_{\text{ebd}}^{\alpha_i\alpha_j}$, 
 and $\text{Net}_{\text{fit}}^{\alpha_i}$), the two scalar factors ($f_{\text{odd}}$ and $f_{\text{dec}}$) together as $\btheta$ and the associated trial wave-function as $\Psi_{\btheta}(\br)$ 
(below for convenience we ignore the dependence on $\bR$, the clamped ion positions).
$\btheta$ are initialized randomly from a Gaussian distribution without any pre-training on pre-calculated wave-functions.
We use VMC to optimize the trial wave-function.
Specifically, we keep track of $N_{\text{wk}}$ walkers to approximate the squared wave-function through an empirical distribution. 
The learning process consists of two phases,  the sampling phase and the optimization phase, which we proceed with alternatively.
In the sampling phase of step $t$, 
we run several steps of the Metropolis-Hasting algorithm to update the positions of the walkers, according to a target probability proportional to $\Psi^2_{\btheta_t}(\br)$. 
In the optimization phase of step $t$, we aim to minimize the second moment of the local energy $E_{\text{loc}}(\br)$ with respect to a fixed reference energy $E_{\text{ref}}$.
Here the local energy is defined as $E_{\text{loc}}(\br)\coloneqq \hat{H}\Psi_{\btheta}(\br)/\Psi_{\btheta}(\br)$. Accordingly, the objective function has the explicit form
\begin{align*}
\Omega_{E_{\text{ref}}}(\btheta) &\coloneqq 
\frac{\int\Psi_{\btheta}^2(\br)(E_{\text{loc}}(\br)-E_{\text{ref}})^2\,d\br}{\int\Psi_{\btheta}^2(\br)\,d\br}.
\end{align*}
In numerical computation, in order to reduce the variance when evaluating the objective function, we use a technique called correlated sampling.
Assuming the walkers' current positions $\{\br^{(1)},\br^{(2)},\dots,\br^{(N_{\text{wk}})}\}$ approximate the distribution $\Psi_{\btheta_t}^2(\br)$ well, we define a reweighting factor $w_i \coloneqq \Psi_{\btheta}(\br^{(i)})/\Psi_{\btheta_t}(\br^{(i)})$ and rewrite the objective function as
\begin{align*}
\Omega_{E_{\text{ref}}}(\btheta)&= 
\frac{\int\Psi_{\btheta_t}^2(\br)(\Psi_{\btheta}^2(\br)/\Psi_{\btheta_t}^2(\br))(E_{\text{loc}}(\br)-E_{\text{ref}})^2\,d\br}{\int\Psi_{\btheta_t}^2(\br)(\Psi_{\btheta}^2(\br)/\Psi_{\btheta_t}^2(\br))\,d\br}\\
&\approx \frac{\sum_{i=1}^{N_{\text{wk}}} w_i^2(E_{\text{loc}}(\br^{(i)})-E_{\text{ref}})^2}{\sum_{i=1}^{N_{\text{wk}}} w_i^2}.
\end{align*}
Note here we take $\btheta_t$ as constants and view both $w_i$ and $E_{\text{loc}}(\br^{(i)})$ as functions of the parameters $\btheta$ given the walker's position $\br^{(i)}$. 
The gradients $\nabla_{\btheta}\Omega_{E_{\text{ref}}}(\btheta)$ are computed by the backpropagation algorithm and used to update the parameters through $\btheta_{t+1}=\btheta_{t} - \eta \nabla_{\btheta}\Omega_{E_{\text{ref}}}(\btheta_t)$, with $\eta$ being the learning rate. 
Motivated by the idea of stochastic gradient descent (SGD), we actually evaluate $\nabla_{\btheta}\Omega_{E_{\text{ref}}}(\btheta)$ with a random batch of data $i\in \mathcal{B}\subseteq \{1,\dots,N_{\text{wk}}\}$ and update $\btheta_t$ with a few steps. 

\section{Results and Discussion}
\begin{table}[!htp]
  \centering
  \caption{Ground-state energy of several systems obtained by DeepWF. 
The bond length of H$_2$, LiH, and H$_{10}$ is 1.4, 1.62, and 1.8 $a.u.$, respectively.
The benchmarks for H$_{2}$, He, LiH, Be, and B are taken from the Computational Chemistry Comparison and Benchmark (CCCBDB) DataBase ~\cite{johnson2018cccbdb}, at the level of configuration interaction, singles and doubles (CISD) theory. 
The benchmark for H$_{10}$ is taken from~\cite{motta2017towards}, at the level of multi-reference configuration with  the Davidson correction (MRCI+Q)~\cite{knowles1988MRCI}. 
All the benchmark results are extrapolated to the complete basis set (CBS) limit.
The relative difference (Rel. Diff) is reported. 
We notice that since we try to use a consistent and accurate theory for the benchmark, the corresponding energies may not be the lowest state-of-art values. For example, in Ref.~\cite{Pekeris1966PR}, C. L. Pekeris et. al. already obtained a value of -2.903724351 $a.u.$ for the ground-state energy of He.}
  \vspace{1em}
  \label{tab:energy}
  \begin{tabular*}{0.78\textwidth}{@{\extracolsep{\fill}}cccc} \hline\hline
    System      & DeepWF~[$a.u.$]     & Benchmark~[$a.u.$] & Rel. Diff \\  \hline
    H${}_2$     & -1.1738    & -1.1741          &   0.26\%     \\
    He          & -2.9036    & -2.9029          &  -0.02\%      \\
    LiH         & -7.8732    & -8.0243          &   1.88\%       \\
    Be          & -14.6141   & -14.6190         &   0.03\%       \\
    B           & -24.2124   & -24.6006         &   1.58\%       \\
    H${}_{10}$  & -5.5685    & -5.6655          &   1.71\%       \\\hline\hline
  \end{tabular*}
\end{table}

\begin{figure}
  \centering
  \includegraphics [width=0.9\textwidth] {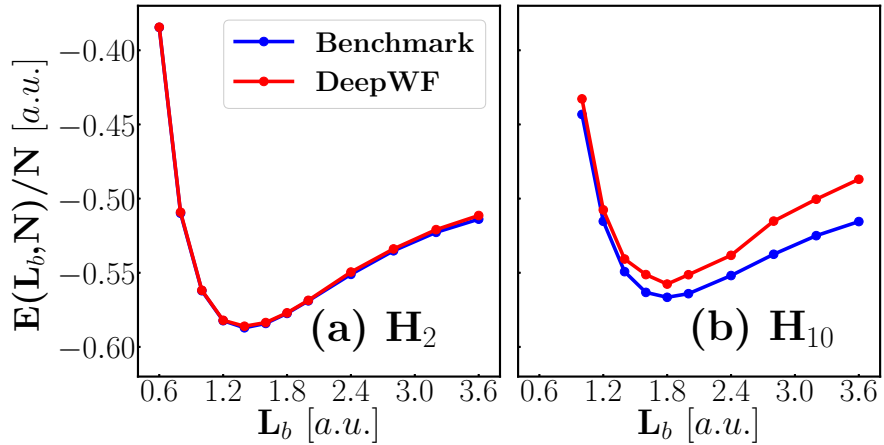}
  \caption{Potential energy per atom $E(L_b,N)/N$ vs. bond length $L_b$ for (a) H${}_2$ ($N=2$), and (b) H$_{10}$ ($N=10$).
  The benchmark result for H$_{2}$ is obtained using the PySCF package~\cite{sun2018pyscf}, at the level of full CI.
 The benchmark result for H$_{10}$ is taken from~\cite{motta2017towards}, at the level of MRCI+Q~\cite{knowles1988MRCI}. 
Both the benchmark results are extrapolated to the CBS limit. }
  \label{fig:eos}
\end{figure}

\begin{figure}
  \centering
  \includegraphics [width=0.9\textwidth] {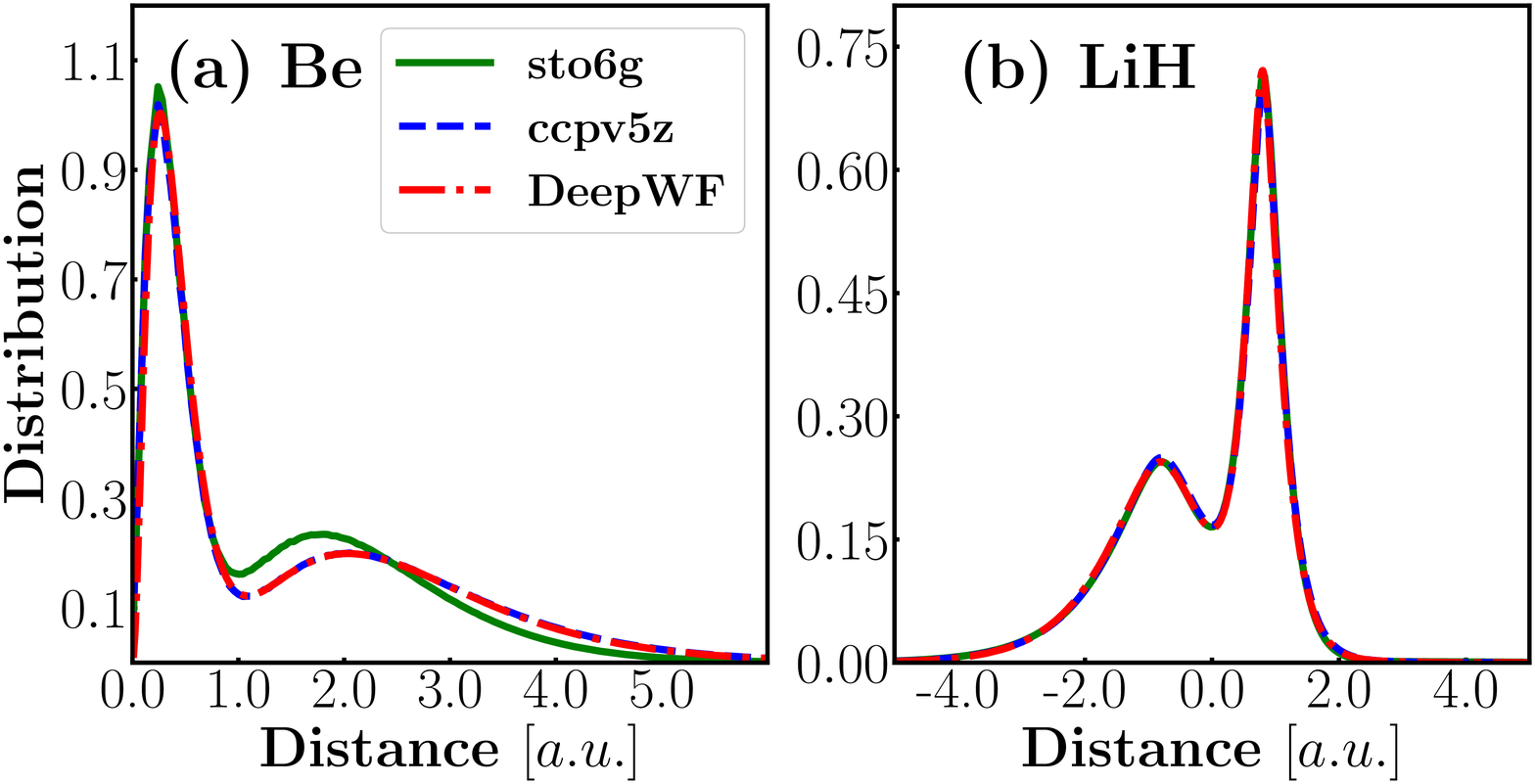}
  \caption{Spatial distribution functions of electrons in different systems. (a) radial distribution of 4 electrons in a Be atom; (b) axial distribution of 4 electrons in a LiH molecule. 
  The results obtained using the PySCF package~\cite{sun2018pyscf} with the sto6g and the ccpv5z bases are plotted for comparison.}
  \label{fig:ele_struc}
\end{figure}

Using the algorithm described above, for a series of benchmark systems, we obtain the optimized parameters $\btheta^{*}$ and run further Monte Carlo simulations for several statistical properties.
The network structure and training scheme are detailed in the appendix.
In Table~\ref{tab:energy}, we present the ground state energies of H${}_2$, He, LiH, Be, B,
and a chain of 10 hydrogen atoms under open boundary conditions.
Note that for H${}_2$ and He we don't have the anti-symmetric part in the wave-function.
Overall the ground-state energies obtained by DeepWF show great consistency with the benchmarks.
However, as the number of electrons increases, the accuracy deteriorates.

In Fig.~\ref{fig:eos}, we plot the potential energy curves of H${}_2$ and H${}_{10}$ as a function of bond length.
The result for H${}_2$ is very close to the benchmark.
For H${}_{10}$, the DeepWF shows a good relative energy, with the correct prediction of local minimum, but is still a bit above the benchmark result.

Finally, Fig.~\ref{fig:ele_struc} plots the spatial distributions of electrons in different systems. 
In the case of radial distribution function of electrons in a Be atom (Fig.~\ref{fig:ele_struc} (a)), it is remarkable that the DeepWF learns from scratch the shell structure of the electrons and shows great consistency with the result obtained using the ccpv5z basis~\cite{woon1995gaussian}.
As a comparison, the sto6g basis~\cite{hehre1969sto} is not enough to describe the electronic dispersion.
In the case of axial distribution function of electrons in a LiH molecule (Fig.~\ref{fig:ele_struc} (b)), DeepWF and the other two methods show excellent agreement.



What is appealing to us is the simplicity of the proposed approach. 
Obviously there is a huge room for further improvement.
In terms of the representation of DeepWF, the symmetric part is relatively general, considering the success of a similar version in representing the inter-atomic potential energy surface. 
However, the anti-symmetric ansatz, although appealing due to its quadratic scaling, might not be sufficient in representing the electronic correlations caused by the Pauli exclusion rule. 
Second, in terms of optimization,
techniques for accelerating VMC through more efficient sampling (see e.g. \cite{lee2011strategies,dewing2000improved}) can be directly adapted into our DeepWF method. 
Optimization method other than the SGD-like correlated sampling method can also be employed.
In any case, we hope that ideas presented here will add some ammunition to the heroic endeavor of attempting to solve the many-body Schr\"odinger equation.


\section*{Acknowledgments}
The authors acknowledge M. Motta for helpful discussions.
This work is supported in part by Major Program of NNSFC under grant 91130005, 
ONR grant N00014-13-1-0338 and NSFC grant U1430237. 
We are grateful for the computing time provided by 
the High-performance Computing Platform of Peking University
and the TIGRESS High Performance Computing Center at Princeton University.

\bibliography{ref}{}
\bibliographystyle{elsarticle-num} 

\newpage
\appendix
\setcounter{table}{0}
\renewcommand\thetable{\Alph{section}.\arabic{table}}
\section{Details of the Training Procedure for Each System}
\begin{table}[!pht]
  \caption{Hyperparameters related to network structure and training scheme for each system.
  }
  \begin{center}
  \label{tab:params}
  \begin{tabular*}{0.98\textwidth}{@{\extracolsep{\fill}}cccccc}
  \hline\hline
   System    & Net$_\text{anti}{}^{\flat}$ & Net$_\text{ebd}^{\alpha_i\alpha_j}$ &
   Net$_\text{fit}^{\alpha_i}$  & LR scheme${}^{\natural}$ & $E_{\text{ref}}{}^{\sharp}$ \\ \hline
   H$_2$     & -          &  (20, 40, 80)  &  (160, 80, 40) &  (5e-4, 1e-8) & -1.3
   \\
   He        & -          &  (20, 40, 80)  &  (160, 80, 40) &  (5e-4, 1e-8) & -3.0
   \\
   LiH       & (40, 40, 40) &  (20, 40, 80)  &  (160, 80, 40) &  (5e-4, 1e-8) & -10.0
   \\
   Be        & (40, 40, 40) &  (20, 40, 80)  &  (160, 80, 40) &  (5e-4, 1e-8) & -16.0
   \\
   B         & (40, 40, 40) &  (20, 40, 80)  &  (160, 80, 40) &  (5e-4, 1e-8) & -26.0
   \\
   H${}_{10}$& (40, 40, 40) &  (20, 40, 80)  &  (160, 80, 40) &  (1e-4, 1e-8) & -6.0
   \\
   \hline\hline
  \end{tabular*}
  \end{center}
  \justify
  \footnotesize{
  $^{\flat}$The network is represented by the number of nodes in each hidden layer, from input to output. The activation function is always hyperbolic tangent, i.e., $\sigma(x)=\tanh(x)$. Net$_\text{anti}$ is not used for two-electron systems H$_2$ and He. The parameters of Net$_\text{anti}^\uparrow$ and Net$_\text{anti}^\downarrow$ are shared except in the case of B, where the number of spin-up and spin-down electrons are different by one. 
Net$_\text{odd}^\uparrow$ and Net$_\text{odd}^\downarrow$ are chosen to be identity functions in this work, since we find no significant effect of network approximation to this term in the tested systems.
  \vspace{0.4em} \\ 
  ${}^{\natural}$We use Adam stochastic gradient descent method~\cite{kingma2015adam} to optimize all the parameters. The first number denotes the starting learning rate and the second denotes the ending learning rate. The learning rate is a exponential decay function respect to the number of epochs, where the decay factor is determined by the starting/ending learning rates and total number of epochs. In all the tested systems, we call 24 batch stochastic gradient descent iterations as an epoch and allow each sample to be used multiple times in an epoch. The batch size and total number of epochs are always 256 and 5000, respectively.
  \vspace{0.4em} \\
  ${}^{\sharp}$The hyperparameters related to VMC sampling are the same for all the tested systems except the reference energy $E_{\text{ref}}$. We use $N_\text{wk}=2048$ walkers and run 80 steps of Metropolis-Hasting sampling each time after one epoch iteration of parameters. The proposal distribution is Gaussian, with the standard deviation being adjusted on the fly such that the acceptance rate stays between 15\% and 75\%.}
\end{table}

\end{document}